\documentclass[referee,usenatbib]{mnras}
\usepackage{mathptmx}
\usepackage[T1]{fontenc}
\usepackage{ae,aecompl}
\usepackage{graphicx}	
\usepackage{amsmath}	
\usepackage{amssymb}	
\title[lens-source degeneracy]{Minimal lensing solutions in the singular perturbative approach.}
\author[Alard, C.]{Alard, C., 
\\
IAP, 98bis Boulevard Arago, Paris \\}
\date{}
\pubyear{2022}
\begin{document}
\label{firstpage}
\pagerange{\pageref{firstpage}--\pageref{lastpage}}
\maketitle
\begin{abstract}
This paper analyse the properties of minimal solutions for the reconstruction of the lens potential in the singular perturbative approach. These minimal
solutions corresponds to an expansion with a minimal degree in Fourier expansion of the perturbative fields. Using these minimal solutions prevent spurious physically meaningless terms in the reconstruction of the fields. In effect a perturbative analysis indicates that a small change in the source model will corresponds to the higher order terms in the expansion of the fields. The results of the perturbative analysis are valid not only for slightly non-circular sources but also for more distorted sources to order two. It is thus of crucial importance to minimize the number of terms used in the modelling of the lens. Another important asset of the minimal solutions is that they offers a de-coupling between the source and lens model and thus help
 to break the source lens degeneracy issue. 
 The possible drawback of minimal solutions is to under-estimate the higher order terms in the solution. However this bias has its merit since the detection of higher order terms using this method will ensure that these terms are real. This type of analysis using minimal solutions will be of particular interest when considering the statistical analysis of a large number of lenses, especially in light of the incoming satellite surveys.
\end{abstract}
\section{Introduction}
\subsection{The general problem of the source shape degeneracy.}
Modeling gravitational lenses does not only require the reconstruction of the lens potential but also the reconstruction of the source. In practice the problem of modeling gravitational lenses is operated in sequence. The first step is to estimate a source model for a given potential model. When the potential is fixed the reconstruction of the source is a linear least-square problem (see \cite{Warren_Dye}). Once a source model has been estimated 
the quality of the reconstruction is estimated by calculating the $\chi^2$ from the difference between the model and the data. Then in the next step the potential is slightly modified in order to search for a better $\chi^2$. This sequence is iterated until a minimum for the $\chi^2$ has been obtained, leading to the best solution for the source
lens system. However it is important to note that in this process there is a complex interaction between the source model and the lens
model. As a consequence it is difficult to understand if the convergence towards a given solution in this process leads to well defined and unique solution, or if the lens and source solution is degenerate. In order to understand the context of the source lens degeneracy problem it is interesting to note the existence of some sophisticated methods of reconstruction. These methods may provide some improvements in the direction of breaking the source lens degeneracy. Some example of these methods are, \cite{Brewer}, \cite{Agnello}, \cite{Saha}, \cite{Meneghetti}. Although these methods are of interest for the source lens degeneracy problem, they do not offer of a full solution to the degeneracy problem. This work will provide a different view of this problem this time based on  the minimal solutions for the lens. The analysis will be based on the singular perturbative approach of gravitational lenses (see \cite{Alard2007}). This paper
will begin with a perturbative analysis of nearly circular sources to prove that a small change in source model generates a large number of higher order
terms in the corresponding model of the perturbative fields. The analysis will be generalized to sources with larger distortion of circularity. Then it will
be shown that minimal solutions are of great interest for breaking source lens degeneracy and that at lower order in the expansion of the expansion
of the perturbative fields the source and lens solutions are almost de-coupled. 
\label{Intro_general}
\subsection{The singular perturbative approach in gravitational lensing.}
In this paper we will work in the singular perturbative approach developed in \cite{Alard2007}, \cite{Alard2008}, \cite{Alard2009}, \cite{Alard2010}, and \cite{Alard2017}. In this approach
the lens equation is simplified and depends only on two functionals of the angular variable $\theta$, the fields $f_1(\theta)$ and $\frac{d f_0}{d \theta}$.
For convenience it is useful to introduce the fields $\tilde f_1(\theta)$ and $\frac{d \tilde f_0}{d \theta}$ which include the additional contributions from
the source impact parameters $(x_0, y_0)$ (see Eq. (\ref{Eq_tilde}), and \cite{Alard2007} Eq. 10). 
\begin{equation}
 \tilde f_i = f_i + x_0 \cos\theta + y_0 \sin\theta ~~~~ {\rm where} ~ i=0,1
\label{Eq_tilde}
\end{equation}
In the singular perturbative approach the source coordinates ${\bf r_s}$ are related to the lens coordinates ${\bf r}$ by Eq. (\ref{Eq_00}). To express
the singular perturbative lens equation it is useful to introduce the parameter $dr$ with $r=1+dr$. Note that for convenience the Einstein radius has been renormalized to unity bu using a proper choice for the coordinates system.
\begin{equation}
 \bold r_s = (\kappa_2 dr -\tilde f_1) \bold u_r - \frac{d \tilde f_0}{d \theta}
\label{Eq_00}
\end{equation}
An interesting application is to consider the image of a round source contour with radius $R_0$ in this theory.
This model is simple with direct analytical formula for the images in singular perturbative expansion (see \cite{Alard2007}, Eq. 12). 
\begin{equation}
 dr=\frac{1}{\kappa_2} \left[\tilde f_1 \pm \sqrt{R_0^2-\left(\frac{d \tilde f_0}{d \theta}\right)^2} \right]
\label{Eq_10}
\end{equation}
The circular contour is transformed into
the images of the source by Eq. (\ref{Eq_10}) (originally Eq. (13) in \cite{Alard2007}).This transformation uses a certain model for the fields $\tilde f_1$ and
$\frac{d \tilde f_0}{d \theta}$. In the forthcoming perturbation theory presented in the next section these fields in the circular source model will be the background fields, 
$f_1^0$ and $d f_0^0$.
\label{Intro_perturb}
\label{Intro}
\section{General perturbation of the round source model.}
\label{sec_1}
In this section we will consider how changing the source model affects the perturbative field solution for the lens.
We will start from the un-perturbed round solution (see Sec. \ref{Intro_perturb}) and its associated fields $\tilde f_1$ and $\frac{d \tilde f_0}{d \theta}$.
Let's consider now that for the same gravitational arc we change the model of the source, which will be now
a perturbed circular source contour. This change in the source contour leads to new equations for the corresponding lens potential.
To the perturbation of the source model a perturbation of the singular perturbative field model will be associated.
Let's define the circular contour source model perturbation $G(x_s,y_s)$,
\begin{equation}
 x_s^2+y_s^2+\epsilon G(x_s,y_s)-R_0^2=0
 \label{Eq_1}
\end{equation}
Here we define the perturbation of the field model when making a model based on the perturbed circular contour defined in Eq. (\ref{Eq_1}).
\begin{equation}
\begin{cases}
\tilde f_1=f_1^0+\epsilon \delta f_1 \\
\frac{d \tilde f_0}{d \theta}=\frac{d f_0^0}{d \theta}+\epsilon \delta f_0
\end{cases}
\label{Eq_1_f}
\end{equation}
The left hand side of Eq. (\ref{Eq_1}) represents the circular source contour, while the right side represents
the perturbation to the circular contour. The corresponding perturbations of the fields in Eq. (\ref{Eq_1}) are $\delta f_1$ and $\delta f_0$.
Here the source coordinates $\bold r_s=(x_s,y_s)$ are related to the lens coordinates and perturbative fields by the perturbative lens
equation (see Eq. \ref{Eq_00}). In Cartesian coordinates Eq. (\ref{Eq_00}) reads,
\begin{equation}
  \begin{cases}
    x_s = (dr -\tilde f_1) \cos \theta +\frac{d \tilde f_0}{d \theta} \sin \theta \\
    y_s = (dr -\tilde f_1) \sin \theta -\frac{d \tilde f_0}{d \theta} \cos \theta
  \end{cases}
 \label{Eq_lens}
\end{equation}
Note that in Eq. (\ref{Eq_lens}) $\kappa_2$ is set to unity. The mass sheet degeneracy implies that $\kappa_2$ is a degenerate
quantity, and for simplicity we set $\kappa_2=1$. It is more convenient to express \ref{Eq_1} in normalized source coordinates
units. We define the following re-scaling,
\begin{equation}
\begin{cases}
x_s \equiv \frac{x_s}{R_0} ~ ; ~ y_s \equiv \frac{y_s}{R_0} \\
\tilde f_i \equiv \frac{\tilde f_i}{R_0} ~ ; ~ f_i^0 \equiv \frac{f_i^0}{R_0}  ~;~ \delta f_i \equiv \frac{\delta f_i}{R_0} ~ ; ~ i=0,1 \\
dr \equiv \frac{dr}{R_0} \\
G(x_s,y_s) \equiv \frac{G(x_S,y_s)}{R_0}
\end{cases}
\label{Eq_2}
\end{equation}
For convenience we will consider that Eq. (\ref{Eq_1}) and Eq. (\ref{Eq_1_f}) are renormalized with the coordinates transformation of Eq. (\ref{Eq_2})
The perturbative solution to the first order in $\epsilon$ for Eq. (\ref{Eq_1}) and Eq. (\ref{Eq_lens}) reads,
\begin{equation}
\begin{cases}
\delta f_1 =\frac{1}{4 \sqrt{1-\left(\frac{d f_0^0}{d \theta}\right)^2}} \left(G\left(\bf {u_r H}, \bf {u_t H} \right)-G\left(\bf {\tilde u_r H}, \bf {\tilde u_t H} \right) \right) \\
\delta f_0 =\frac{1}{4 \frac{d f_0^0}{d \theta}} \left(G\left(\bf {u_r H}, \bf {u_t H} \right)+G\left(\bf {\tilde u_r H}, \bf {\tilde u_t H} \right) \right) 
\end{cases}
\label{Eq_4}
\end{equation}
With the definition of the vectors:
\begin{equation}
\begin{cases}
{\bf u_r}=(\cos \theta, \sin \theta) \\
{\bf u_t}=(-\sin \theta, \cos \theta) \\
{\bf \tilde u_r}=(-\cos \theta, \sin \theta) \\
{\bf \tilde u_t}=(\sin \theta, \cos \theta) \\
{\bf H}=\left(\sqrt{1-\left(\frac{d f_0^0}{d \theta}\right)^2}, \frac{d f_0^0}{d \theta} \right)
\end{cases}
\end{equation}
To illustrate the nature of the perturbative solution presented in Eq. (\ref{Eq_4}) it is interesting to consider
simple models of perturbation of the circular contour. One model of particular interest is the case where the
function $G$ is a polynomial function of the coordinate of order $n$. The lowest order of interest is $n=2$,
the polynomial terms of order two are, $x_S^2$, $x_S y_S$, $y_S^2$. As an example, the perturbation associated with the $x_S^2$ term is,
\begin{equation}
  \begin{cases}
    G(x_s,y_s)=x_S^2 \\
    \delta f_1=-\frac{1}{2} \frac{\tilde d f_{0}}{d \theta} \sin 2 \theta \\
    \delta f_0= \left(\frac{d f_0^0}{d \theta}\right)^{-1}\frac{1}{4} \left(\cos 2 \theta+ 1 \right) - \frac{1}{2} \left(\frac{d f_0^0}{d \theta}\right) \cos 2 \theta
  \end{cases}
\label{Eq_5}
\end{equation}
In the Fourier series expansion of the fields it is apparent in Eq. (\ref{Eq_5}) that the leading
term is, $\frac{d f_0^0}{d \theta} \sin \left(2 (\theta+\zeta)\right)$, with $\zeta=0$ or $\frac{\pi}{2}$.
An elliptical model aligned with the coordinates system axis corresponds to $G(x_S,y_S)=\eta \left(y_S^2 -x_S^2 \right) $. For this particular elliptical model  we retrieve the equation for the elliptical contour in (\cite{Alard2007}) when an
expansion in first order in $\eta$ is performed. In the same way an order $n$ polynomial in $(x_S,y_S)$ the leading term $Q$ in the
Fourier series expansion of the fields is,
\begin{equation}
  \begin{cases}
    G(x_s,y_s)=x_S^m y_S^{n-m}\\
    Q_n \propto  \left(\frac{d f_0^0}{d \theta}\right)^{n-1} \cos \left( n \theta + \zeta\right) ~ {\rm with} ~ \zeta=0,\frac{\pi}{2}
  \end{cases}
\label{Eq_55}
\end{equation}
We define the order of the Fourier series expansion of the unperturbed field $\frac{d f_0^0}{d \theta}$, $n_0$.
Then it is interesting to note that the order $N_F$ of the Fourier expansion according to Eq. (\ref{Eq_55}) is,
\begin{equation}
    N_F=(n-1) n_0 +n
\label{Eq_5.1}
\end{equation}
 It is clear that from Eq. (~\ref{Eq_5.1}) for $n \ge 2$, which is the lowest order of source perturbation (ellipticity) the order of $Q_n$ is greater than $n_0$. This implies clearly that for $n \ge 2$ the perturbative expansion Fourier series order is larger that the order of the original unperturbed term. The obvious consequence is that if the solution for the source is a circular contour, considering a non-circular contour with
order beyond $n=2$ (elliptical) will induce higher order terms in the solution for the fields $\tilde f_1$ and $\frac{d \tilde f_0}{d \theta}$. Note that the order
of the Fourier expansion increase very quickly. Even for the lowest order source perturbation ($n=2$), and for the lowest order lens model, ($n_0=2$), we go from
order 2 to order 4 in the Fourier expansion of the fields which is a very significant increase (see Eq. ~\ref{Eq_5.1}). Considering perturbation of the source of higher order will only increase further the higher order terms if the expansion of the fields.
It is also true that if the best solution is some perturbed non-circular contour, a solution for the fields with a circular contour will contain higher order terms. This is simply because an elliptical contour solution is a circular contour
solution plus a higher order perturbation. Let $S_1$ be the true, most minimal non circular solution, $C_0$ the circular solution, and $T_1$ the higher order terms associated with the first order expansion. Since $S_1$ is a minimal non-circular solution, then due to Eq. (\ref{Eq_C}) the circular solution
will contain additional higher order terms.
\begin{equation}
  S_1=C_0+T_1 \Rightarrow C_0=S_1-T_1
 \label{Eq_C}
\end{equation}
Similarly if we were trying to make a model of a non-circular contour with a different non-circular contour the effect would be additional higher order terms in the Fourier series of the fields. Here
this result is simply the consequence that a non-circular solution is a circular solution plus additional higher order terms,
and that the difference between two non-circular solutions is the difference between two circular perturbed models. In similarity to Eq. (\ref{Eq_C})
if $S_2$ is a non-circular, non-optimal solution with associated terms $T_2$ we obtain,
\begin{equation}
  S_2=S_1+T_2-T_1
 \label{Eq_C2}
\end{equation}
%
%
It is clear from Eq. (\ref{Eq_C2}) that the solution $S_2$ contains the additional higher order terms $T_2-T_1$ with respect to the optimal solution $S_1$.
As a consequence for a source model which is a general perturbation to the first order of the circular source model, the proper solution is the solution with the minimum degree expansion of the fields.
\section{Second order expansion}
\label{sec_2}
In this section we will demonstrate that the results obtained in the former section are still valid not only for nearly round models
but also for more distorted models. To prove this result we will study more elongated contours by pushing the expansion to order 2.
We modify Eq. (~\ref{Eq_1_f}) by expanding the perturbative fields to order 2, leading to Eq (~\ref{Eq_2_f}).
\begin{equation}
\begin{cases}
\tilde f_1=f_1^0+\epsilon \delta f_1+\epsilon^2 \delta f_{1,2} \\
\frac{d \tilde f_0}{d \theta}=\frac{d f_0^0}{d \theta}+\epsilon \delta f_0+\epsilon^2 \delta f_{0,2}
\end{cases}
\label{Eq_2_f}
\end{equation}
By using Eq's (~\ref{Eq_1}), (~\ref{Eq_lens}), and (~\ref{Eq_2_f}) it is possible to calculate the perturbative solution to order 2. The calculations are heavier
than for the first order expansion and we will analyse the solution main properties and behavior at higher order. The main result is that the second order terms,
$ \delta f_{1,2}$ and $ \delta f_{0,2}$, are of higher order than the first order terms except when the perturbation to the contour is of the lowest order ($n=2$). For
polynomial functionals of order $n$ corresponding to $G(x_x,y_s)=x_s^m y_s^{n-m}$, and for $n > 2$ the second order terms are of higher order than the first order terms. To illustrate the structure of the second order terms we express the solution for $n = 2$,
\begin{equation}
  \begin{cases}
    G(x_s,y_s)=x_S^2 \\
    \delta f_{1,2}=-\frac{d f_0^0}{d \theta}^{-1}\left[ \left(\frac{1}{4} \frac{d f_0^0}{d \theta}^2+\frac{1}{8} \right) \sin 2\theta+\frac{1}{16} \sin 4 \theta \right] \\
    \delta f_{0,2}=\frac{d f_0^0}{d \theta}^{-3}\left[\left( -\frac{1}{64} \right) \cos 4\theta+\left(-\frac{1}{16}+\frac{1}{8}\frac{d f_0^0}{d \theta}^2-\frac{1}{4}\frac{d f_0^0}{d \theta}^4 \right) \cos 2\theta -\frac{3}{64}+\frac{1}{8}\frac{d f_0^0}{d \theta}^2+\frac{1}{8}\frac{d f_0^0}{d \theta}^4 \right]
  \end{cases}
\end{equation}
For $n>2$ the leading higher order therm $Q_n$ is,
\begin{equation}
 Q_n=\cos(2 n \theta+\zeta) \frac{d f_0^0}{d \theta}^{2n-3}
\end{equation}
As a consequence the expansion at order two is at least of the order of the first order expansion. This result demonstrates the existence of higher order terms for more elongated sources and generalized the result obtained for first order small distortions of nearly circular sources.
%
\section{Considering minimal solutions help breaking the lens source degeneracy.}
\label{Sec_dec}
\begin{figure}
\includegraphics[width=14cm]{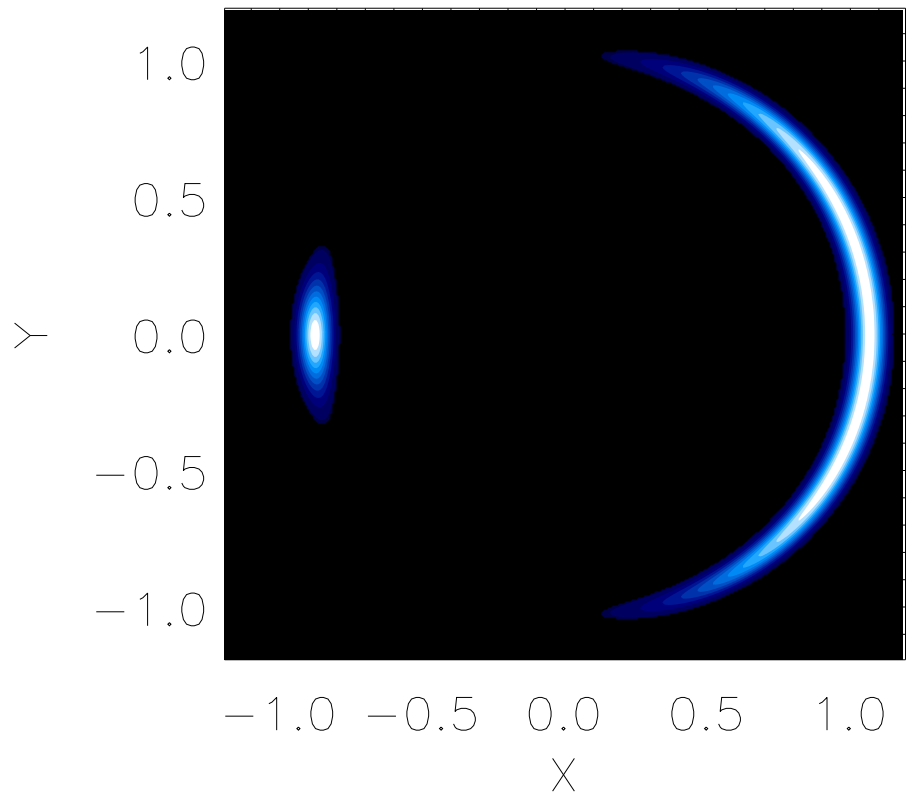}
\caption{The contours of the image of the arc used for the numerical experiments. This image corresponds to a cusp configuration for a an elliptical
isothermal potential. This arc is the image of an elliptical source. This image is represented on a well sampled grid with no noise, it is designed
to reconstruct the source with maximum accuracy, it is not intended to represent real conditions. The image presented if Fig. (\ref{Fig_6}) is more
representative of real conditions.}
\label{Fig_1}
\end{figure}

\begin{figure}
\includegraphics[width=14cm]{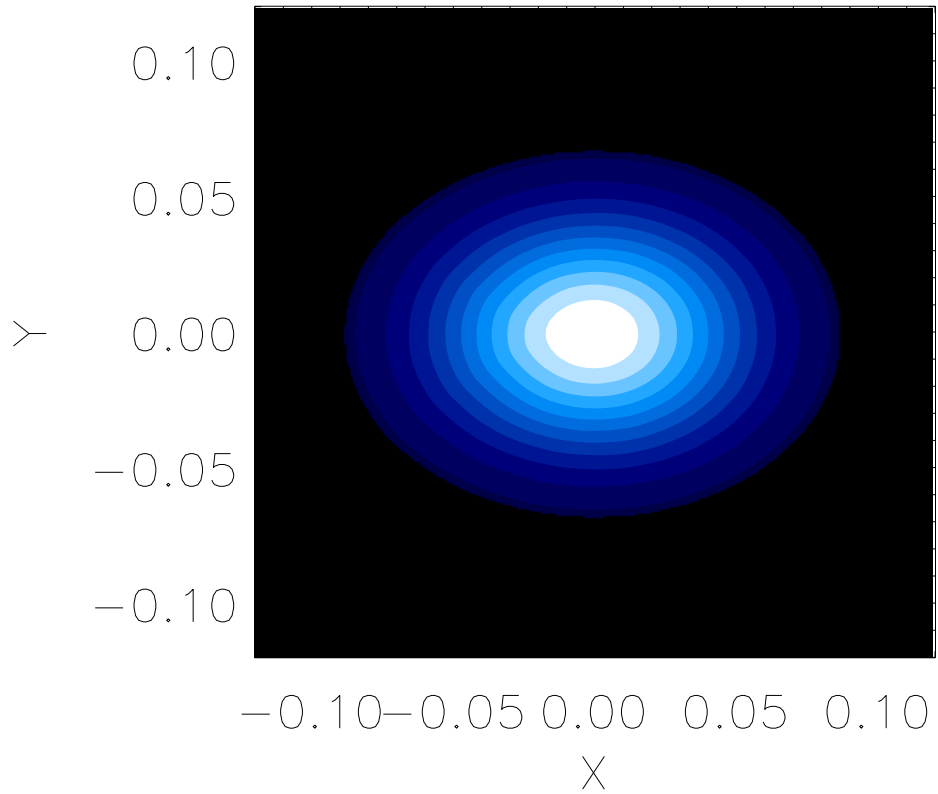}
\caption{The reconstruction of the source corresponding to the arc of Fig. (\ref{Fig_1}). The source is reconstructed by re-mapping the lens coordinates
to the source coordinates using the lens equation. The potential used in the lens equation corresponds to the true lens potential to order 2 in the Fourier expansion of the fields.}
\label{Fig_2}
\end{figure}

\begin{figure}
\includegraphics[width=14cm]{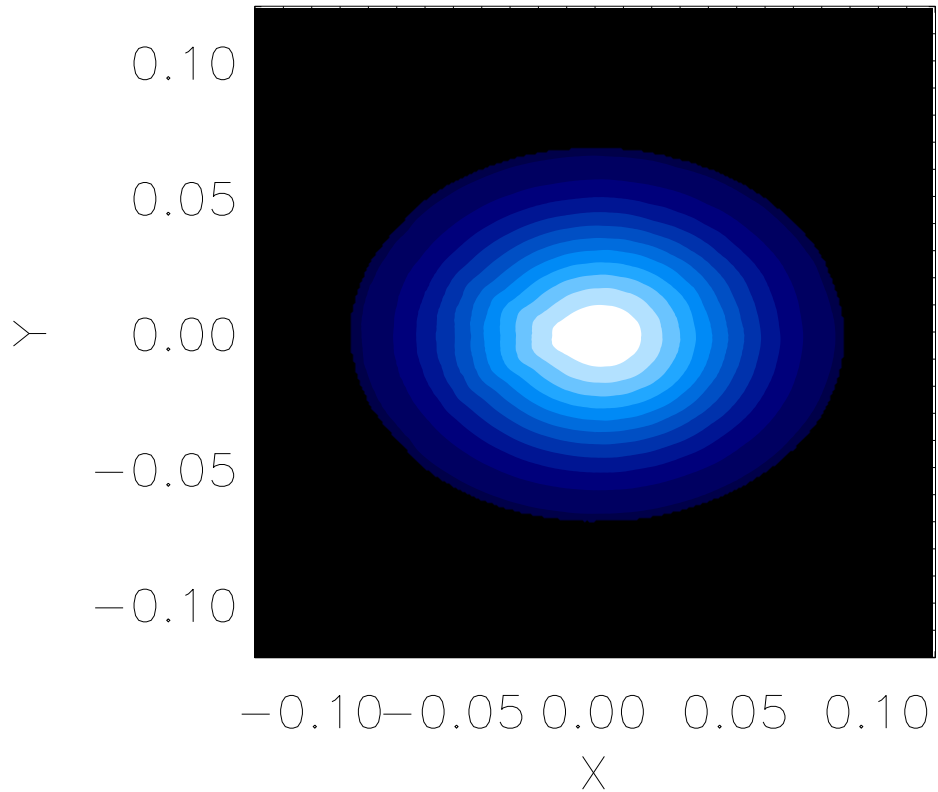}
\caption{Same reconstruction as Fig. (\ref{Fig_2}) but this time the potential used in the lens equation is estimated by fitting a circular source model
to the data presented in Fig. (\ref{Fig_1}). The circular source model is adjusted by using a minimization procedure in the parameter space. The minimization method is based of the Amoeba method. The starting guess for the minimization is the true potential. Some experiments were performed by adding random
fluctuation of the initial guess. These experiments show that the convergence to the solution is stable and does not depend on the initial guess.}
\label{Fig_3}
\end{figure}

\begin{figure}
\includegraphics[width=14cm]{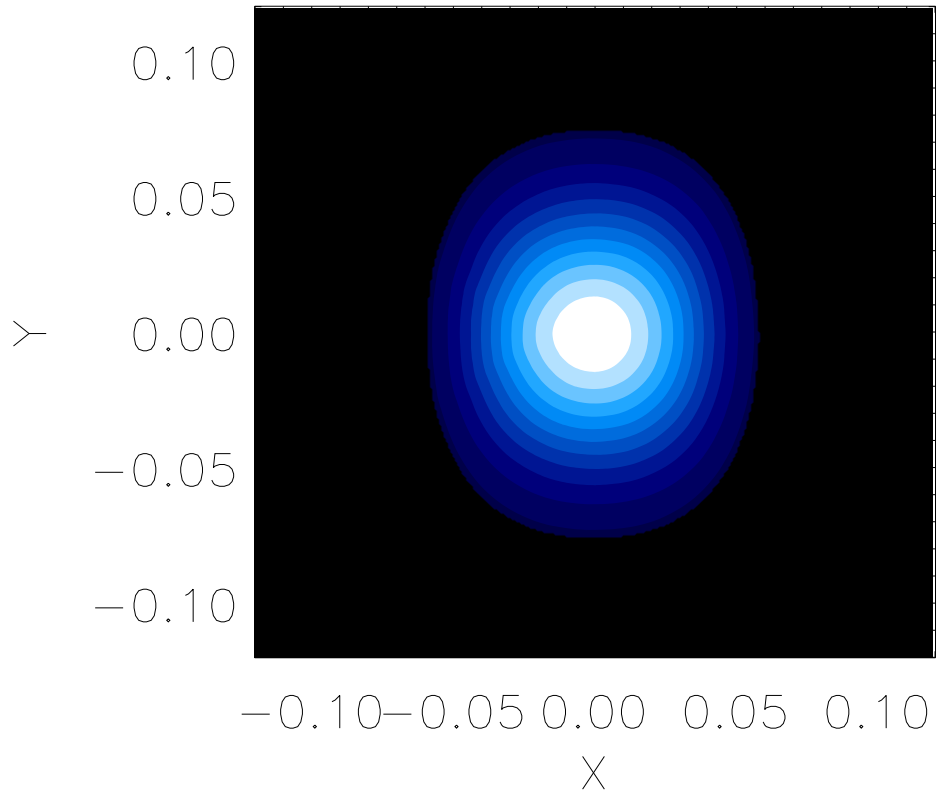}
\caption{Same as Fig. (\ref{Fig_2}) except for a change in the model of the source. In this case the source contour is distorted by a fourth order term (See Eq. (\ref{Eq_G4})).}
\label{Fig_4}
\end{figure}

\begin{figure}
\includegraphics[width=14cm]{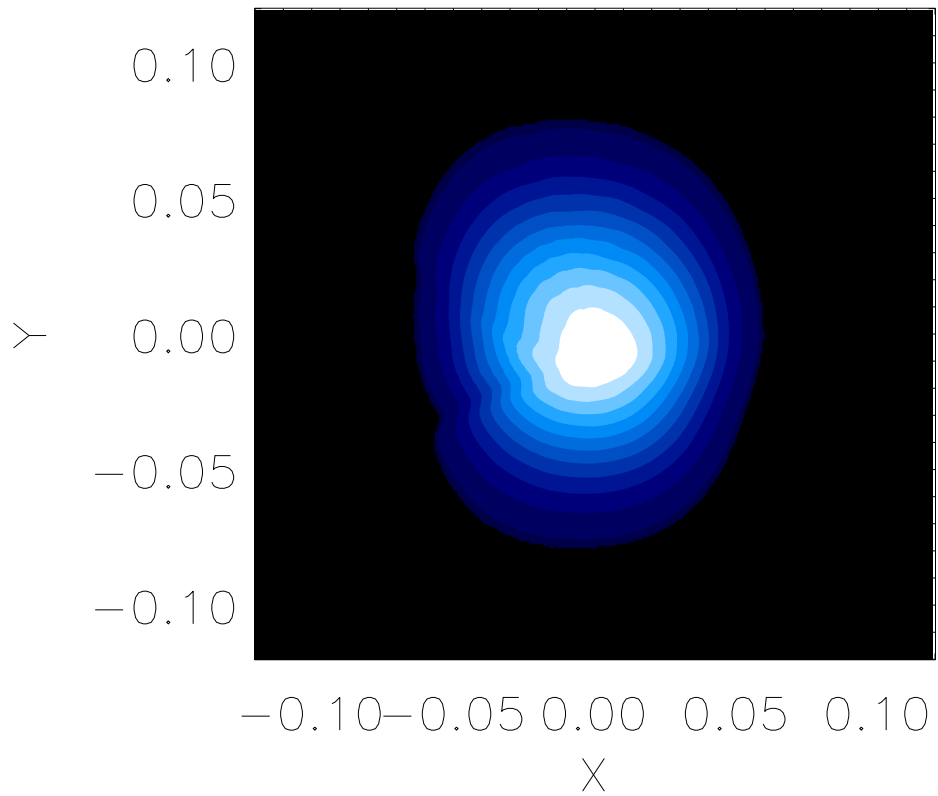}
\caption{Same as Fig. (\ref{Fig_3}) except for a different source model. The source model is defined in Eq. (\ref{Eq_G4}). }
\label{Fig_5}
\end{figure}

\begin{figure}[tb]
\includegraphics[width=14cm]{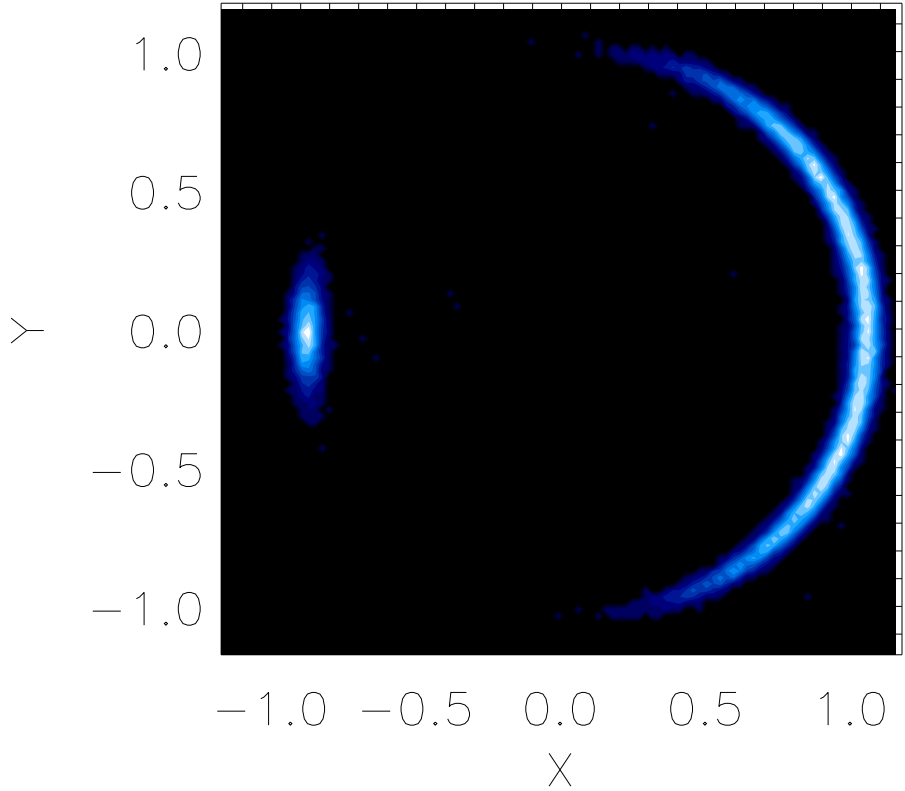}
\caption{The configuration described in Fig. (\ref{Fig_1}) represented in a coarser grid with the addition of Poisson noise. The under-sampling and noise
of this new image are more representative of real data that Fig. (\ref{Fig_1}).}
\label{Fig_6}
\end{figure}

\begin{figure}
\includegraphics[width=14cm]{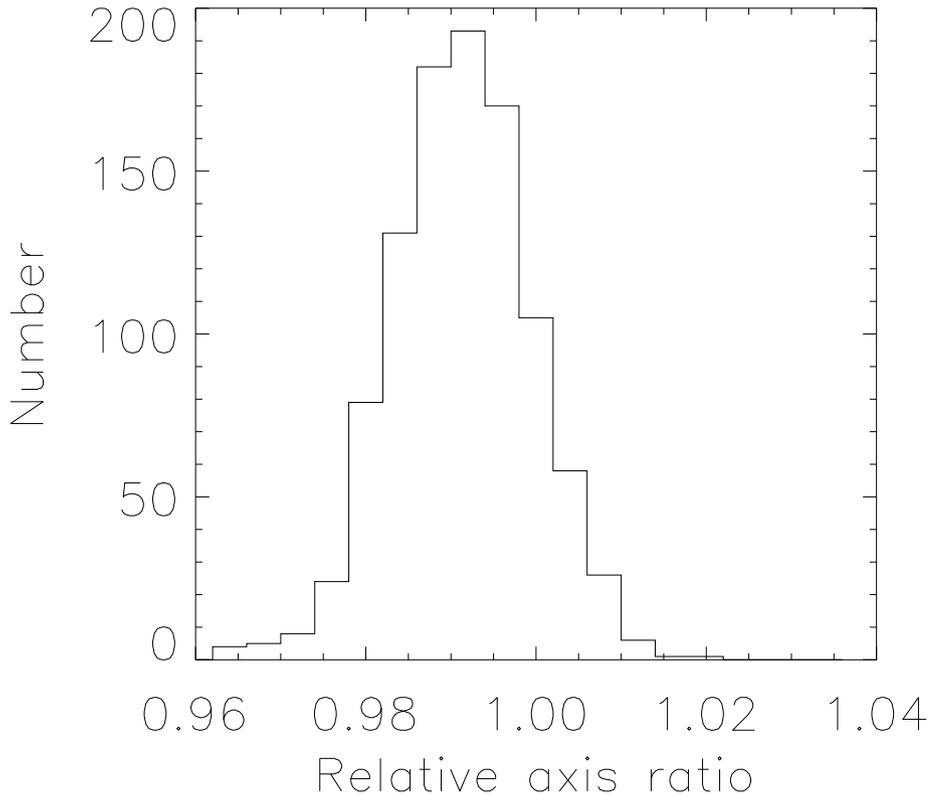}
\caption{Histogram of the relative axis ratio for 1000 simulations for the under-sample arc with added Poisson noise presented in Fig. (\ref{Fig_6}). The relative axis ratio corresponds to the ratio between thes estimated axis ratio and the true axis ratio. For each simulation the source is reconstructed with a potential obtained by fitting a circular source model. The fitting procedure is identical to the procedure used in the former numerical experiments. The initial guess for the fitting procedure is different for each simulation. This initial guess is obtained by adding a random fluctuation to the true potential for each of the Fourier coefficients of the expansion of the perturbative fields. The amplitude of the random fluctuation is a fraction of the mean amplitude of the Fourier coefficients of the true potential Fourier expansion.}
\label{Fig_7}
\end{figure}

If we consider a gravitational arc which is the image of an elliptical source it is definitely possible to make a model of this arc with a circular source.
However as we have seen in the former sections (see Sec. \ref{sec_1} \& Sec. \ref{sec_2}) a change in the source model will produce a number of higher order
terms in the expansion of the fields. As a consequence if we make a minimal model of the arc we should retrieve the elliptical model and not the circular
one which would require many higher order Fourier terms in the expansion of the fields. This is confirmed by analyzing the singular perturbative solution for an elliptical source either from Eq. (15) in \cite{Alard2007} or here in Eq. (\ref{Eq_4}). In effect by developing to the first order in the ellipticity of the source we find that there is very little effect of the ellipticity of the source on the expansion of the fields. For instance if we make a model of the unperturbed circular model with a Fourier expansion at order 3 we find that to the first order there is no terms corresponding to the ellipticity of the source in the perturbed fields. The coupling between the ellipticity of the source and the field expansion occur only when the potential has effectively
significant fourth order Fourier terms. However it is important to note that in practical cases the higher order terms in the expansion of the potential
have quickly decreasing amplitude (see \cite{Alard2009} \cite{Alard2010} \cite{Alard2017}) and that as a result the coupling of the field expansion with source ellipticity should be small. These results are illustrated by taking a simple lens model, an elliptical isothermal lens potential and a cusp configuration (see Fig. \ref{Fig_1}).Basically the elliptical isothermal potential we choose is well described by a second order Fourier expansion in the singular perturbative model.
This potential has only very weak terms beyond order 2. We then make the following experiments, first we reconstruct the source by transforming the arc to the source plane coordinates by using the true lens potential (see Fig. \ref{Fig_2}), and second we reconstruct the lens potential by fitting a circular source model and transform to source coordinates using this second potential (see Fig. \ref{Fig_3}). The potential that we fit in the second reconstruction with a circular source model is minimal which means that the expansion is limited to order 2 . It is clear from Fig's \ref{Fig_2} and \ref{Fig_3}, that
the difference between the two reconstructions are very weak, which proves that there is very little coupling between the source model and the potential.
The difference in the axis ratio between the two source reconstructions is less than one percent of the ellipticity. The difference between Fig's \ref{Fig_2} and \ref{Fig_3} are due to weak higher order terms in the expansion
of the fields which were not taken into account in the expansion which is limited to order 2. The same analysis can be conducted for higher order source
distortion. The same calculations were performed by replacing the elliptical source model by a source distorted by fourth order quartic terms. We choose 
the following model for the source (see Eq. \ref{Eq_1}),
\begin{equation}
 G(x_s,y_s) \propto x_s^4
\label{eq_G4}
\end{equation}
With the source model of Eq. \ref{Eq_G4} as before we reconstruct the source by transforming to source coordinates using the true potential (see \ref{Fig_4}), and for comparison we reconstruct the lens potential to order 2 using a circular source model and use this potential to reconstruct the source
by moving to source coordinates (see Fig. \ref{Fig_5}). This time we observe some difference between the two reconstructions of the source. However it is important to note that the differences are not large and that the general properties of the source are close in each case. For instance the difference between the coefficients of the fourth order moment of the source in each case is of only of about five percent. This is larger than what we observed for an elliptical source but this is still quite close which shows that the coupling is not very large between the source model and the field reconstruction even 
for higher order source distortion. This discussion shows that in general we expect that the coupling between the source model and the potential model
will be quite small. As a consequence it is clear that using an iterative method where the first step is to make a circular source model estimate the potential and then estimate the source, and then using the improved source model and modeling the potential and so on will lead to a quick convergence towards the right source lens solution. Provided of course that the potential model is minimal and does not include un-necessary spurious higher order terms. Here we see all the interest of the minimal solution for the perturbative fields. The numerical experiments we described are designed to analyze
the effect of potential modeling on the source reconstruction with maximum accuracy, and as a consequence do not include noise in the image of the gravitational arc. It is interesting to make new numerical experiments with realistic noise and sampling to see if the results obtained with the former numerical experiments still holds in the presence of noise. To investigate this issue the same cusp configuration was simulated but this time with Poisson noise and some under-sampling in order to be consistent with real data (see Fig. \ref{Fig_6}). For this simulation process an elliptical source was used.
The simulation process include a total of 1000 simulations. For each simulation a new image image is simulated with a given expectation for the Poisson noise. Then a circular source model with an order 2 potential is fitted to the data. As before the fitting processes is performed in the multi parameter space using a minimization method based on the amoeba method. To initiate the parameter search the guess is the true solution for the fields plus a random fluctuation of the parameters which is of the order of a fraction of the mean value of the parameters. The fluctuation in the guess is performed to analyze the sensitivity
of the minimization process to the guess. The results presented in Fig. (\ref{Fig_7}) shows the stability of the solution and the general convergence of the minimization process to a small area of the parameter space. In  Fig. (\ref{Fig_7}) the stability of the solution and the general error is analyzed by evaluating the axis ration of the reconstructed source. A small bias of the order of 1 \% is observed in the mean value of the axis ratio, while the scatter
is of about the same order. The result of these simulations show that once again the source and potential model are de-coupled even in the presence of noise,
provided that a minimal model is used for the perturbative fields. 
\clearpage
%
%
%
\section{Conclusion.}
In this paper new approach to the lens source degeneracy is proposed by considering minimal lens solutions. In the perturbative approach
the minimal solution corresponds to the minimum degree in the Fourier expansion of the perturbative fields. In the modeling of gravitational lenses it is essential to consider these minimal solutions. The minimal solutions are a well defined fixed point in the space of solutions. 
Trying to represent the potential of gravitational lenses with unnecessary higher order expansions leads to a quickly increasing number
of spurious higher order terms. This is due to the fact that in response to a small change of the source model from the model corresponding
to the optimal solution the perturbative response in the lens solution grows very quickly,  leading to a  high number of higher order terms.
These higher order terms do not have any physical meaning or any value in the interpretation of the lens model. Another very important asset of the minimal 
solutions is that in general they offer some de-coupling between the source and lens solution. As discussed in Sec. (\ref{Sec_dec}) the terms associated with the distortion from circularity in the source model have only a weak influence in the perturbative field model provided that the model for the fields is minimal. In Sec. (\ref{Sec_dec}) a simple model with dominant order 2 term was explored. 
We may ask now what would be the situation for a more complex lens model ? This problem was discussed in Sec. \ref{Sec_dec}) and the idea is that
even if the lens model contains significant terms beyond order 2 in practice these terms will be of decreasing amplitude and thus should not have a major influence on the lens source coupling. The only case where higher order Fourier terms can be large is when the potential is perturbed locally with for instance
the presence of a substructure (see \cite{Alard2008}). But it is clear that in the case of a local structure the higher order terms are significant only
in the vicinity of the substructure, and that most of the arc will not require these high amplitude higher order terms for its modeling. As a consequence we expect that even in this case the lens source problem will be de-coupled to some degree. 
The problem now is to discuss what is a minimal model for a given gravitational arc. This minimal model should be explored carefully and individually for each gravitational arc system. It is of crucial importance to look for solutions with the least possible number of terms in order to avoid introducing spurious terms in the solution. When reconstructing the solution it is essential to ensure that at the lowest order the best minimization has been achieved before any additional higher order terms are introduced. At least in the beginning of the process it is essential to explore the most minimal solution
 possible in order to ensure a proper de-coupling of the source and the lens. 
 This means that the minimization process and search for the best solution at the lowest order for the potential has to explore thoroughly the parameter space and ensure that the absolute minima for a given order of the Fourier expansion has been found.
 Once a convergence has been obtained for the most minimal model, only then
 can the fitting procedure can be extended to include higher order terms. It is clear also that the search for higher order terms should be done
 in the vicinity of the most minimal former solution to avoid reaching a spurious higher order solution. The additional higher order terms
 should be minimal or have the least possible amplitude. A possible drawback of enforcing the minimal solution is that
the solution may be biased and minimize the higher order terms. But this drawback is also an advantage since in the minimal approach if higher order terms are found in the solution it means that these terms are real and not spurious oscillations of the solution. This point is of special importance if we consider the problem of substructure in dark halos. Since the presence of substructures is related to the existence of of higher order terms in the solution (see Alard ~\cite{Alard2009}), it is of special importance to look for the minimal solution. A statistical analysis based on minimal solutions will ensure that the higher order signal present in the statistics of the solutions is real and an effect of spurious terms. The same is true for other manifestations of higher order
terms like the complex geometry of the dark matter halos which as a consequence may not follow the distribution of light (see ~\cite{Alard2010}). As a conclusion it is clear that looking for the minimal solution should be the standard procedure in the reconstruction of gravitational lenses in the singular perturbative approach. To conclude it is clear that the minimal solutions have very interesting and promising properties for the reconstruction of gravitational lenses. However more work is required to explore the properties of these solutions in practical applications to evaluate how the minimality of the solution can be enforced for various lens systems. The problem of the practical application of the method will be explored in more detail in a forthcoming paper.
\section*{Data Availability}
No data sets were generated or analyzed during the current study.
\end{document}